\documentstyle[twocolumn,aps,epsf,psfig]{revtex}
\begin{document}
\draft

\twocolumn[\hsize\textwidth\columnwidth\hsize\csname @twocolumnfalse\endcsname

\title{Revisiting two holes in a locally antiferromagnetic 
background:\\ the role of retardation and Coulomb repulsion effects}

\author{Jos\'e Riera$^1$ and Elbio Dagotto$^2$}
\address{$^1$ Instituto de F\'{\i}sica Rosario y Departamento de
F\'{\i}sica,  \\ Avenida Pellegrini 250, 
2000 Rosario, Argentina}
\address{$^2$ Department of Physics and National High Magnetic Field Lab, 
Florida State\\
University, Tallahassee, Florida 32306, USA}
\maketitle

\begin{abstract}

The problem of two holes in the presence of strong antiferromagnetic
fluctuations is revisited using computational techniques. Two-dimensional
clusters and 2-leg ladders are studied with the Lanczos  and Truncated
Lanczos algorithms. Lattices with up to $2 \times 16$ and 
$\sqrt{32} \times \sqrt{32}$ sites are studied. The
motivation of the paper is the recently discussed spatial distribution
of holes in ladders
where the maximum probability for the  hole-hole distance  
is obtained at $d = \sqrt{2}$ in
units of the lattice spacing, a counter-intuitive result considering that
the overall symmetry of the  two-hole bound state is $d_{x^2 - y^2}$.
Here this effect is shown to appear in small ladder clusters 
that can be addressed exactly, and also in planes.
The probability distribution of hole distances $d$ was found to be broad
with several distances contributing appreciably to the wave function.
The existence of holes in the same sublattice is argued to be
a consequence of non-negligible retardation effects in the $t-J$ model.
Effective models with instantaneous interactions nevertheless
 capture the essence of
the hole pairing process in the presence of short-range antiferromagnetic
fluctuations (specially regarding the symmetry properties of the 
condensate), similarly as the (non-retarded)
BCS model contains the basic features of the more complicated
electron-phonon problem in low temperature superconductors. 
 The existence of strong
spin singlets in the region where the two hole bound state is
located is here confirmed, and a simple explanation for its origin
in the case of planes is
proposed using the N\'eel state as a background, complementing previous
explanations based on a spin liquid undoped state. 
It is predicted that these strong singlets should appear
regardless of the long distance properties of the spin system under 
consideration, as long as the bound state is $d_{x^2 - y^2}$. 
In particular, it is shown that they are present in an Ising spin 
background.  The time retardation
in the family of $t-J$ models leads naturally
 to low-energy hole states with nonzero momentum
and spin one, providing a possible explanation for apparent 
SO(5)-symmetric features observed recently in this context.
Finally, the influence of a short-range Coulombic
repulsion is analyzed. Rough estimations suggest that at a distance of
one lattice spacing this repulsion is larger than the exchange $J$. The
hole distribution in the $d_{x^2 - y^2}$ bound state is reanalyzed in
the presence of such repulsion. Very short hole-hole
distances lose their relevance in the presence of a realistic
hole-hole interaction.

\end{abstract}


\vskip2pc]
\narrowtext

\section{Introduction}

Among the most appealing scenarios to explain the behavior of the
high temperature superconducting compounds are those based on 
antiferromagnetic fluctuations as the mechanism for hole pairing.
These ideas have been formulated over the years in a variety of 
contexts. Early diagrammatic work for heavy fermions in the presence of
antiferromagnetic
correlations suggested that magnon interchange can lead to $d$-wave
pairing~\cite{doug1}. The nearly antiferromagnetic Fermi liquid proposal 
arrives to the same conclusion also in a
diagrammatic context where low-energy spin excitations are assumed to
coexists with holes~\cite{pines}. Since
in real doped cuprates the antiferromagnetic
correlation length $\xi_{AF}$ is only of a few lattice spacings other
approaches have emphasized the relevance of the
short distance physics in the problem~\cite{arno}.
Independently of these approaches which are based on the 
resummation of particular sets of Feynman diagrams,
a vast amount of computational work has been devoted to the interaction
of holes with spin excitations~\cite{review}. 
Although these techniques have as a natural
limitation the finite size of the clusters that can be studied, they
have the important
advantage that inside these clusters the information gathered is
basically exact. Since $\xi_{AF}$ is small in the cuprates,
and, in addition, experimental estimations of the
coherence length also suggest small size Cooper pairs~\cite{cyrot},
computational approaches are very important to clarify the physics of
carriers in a fluctuating antiferromagnetic background.
Following these ideas  the existence
 of pairing of two holes  in the $d_{x^2 - y^2}$ channel using 
the $t-J$ and Hubbard models was
observed since the early theoretical studies of the
cuprates~\cite{review,doug2}. For 
realistic values of $J/t$ the size of the pair has been known to be
small, of only a couple of lattice spacings at most~\cite{didier1}. In
addition, recent work in
the limit where the attraction between holes is assumed much larger
than the bandwidth provided a real-space simple qualitative picture
for the formation of $d_{x^2 - y^2}$ 
bound states~\cite{naza}. It turns out that
not only a suitable effective potential between holes is needed to 
obtain pairs in the $d$-channel
(repulsive on site, and attractive between nearest-neighbors) but
in addition the form of the dispersion of holes renormalized by spin
fluctuations plays an essential role in stabilizing $d$-wave pairing
instead of extended $s$-wave pairing~\cite{naza}.

The small tight pairs found numerically suggest that the cuprates are 
in an intermediate regime between
the BCS limit, where the pair size is much larger than the mean distance
between carriers, and the Bose condensation regime defined in the
opposite limit. Recent experimental work reporting the presence of
pseudogap features in the normal state of the underdoped
cuprates~\cite{pes} have  revived the
possibility that hole pairs may exist above the superconducting
critical temperature, at least as finite lifetime 
fluctuations~\cite{prefor}.
These results provide extra support to the idea that a ``real-space'' 
approach to hole pair formation is more enlightening than a 
momentum-space approach.
In the context of holes in antiferromagnets
it is natural to propose an effective model for holes
where these particles move within the same sublattice (as suggested by
a variety of studies of the one hole problem in an antiferromagnet),
interacting through an effective potential induced by
antiferromagnetism~\cite{afvh}.
In the limit where this potential is assumed to be short-ranged the
presence of $d$-wave bound states has been deduced analytically through 
the solution of the two-body problem, and the
presence of superconducting correlations has been found using
the BCS mean-field approach with RPA fluctuations supplemented by 
numerical studies~\cite{naza}.
More recent work has extended these ideas to include the
interchange of magnons, leading to a longer range effective potential
between holes reaching similar conclusions regarding the $d$-wave pair
formation in the problem~\cite{dotsenko}.

Recently, a complementary approach to the study of holes in
the cuprates has been considered. It is based on the analysis of
even-leg ladder systems~\cite{ladder} since they have a robust spin 
gap, a feature that is also
present in the normal state of underdoped cuprates. The 2-leg ladder
has an antiferromagnetic correlation length of about 3 lattice
spacings~\cite{doug3}, similar to the correlation observed in the doped
cuprates. Then, ladders have coexisting spin gap features and short-range
antiferromagnetism.
In this
context it has been observed that a couple of holes introduced in
the 2-leg ladder form a bound state for realistic values of $J/t$ 
with characteristics of a $d$-wave pair~\cite{sigrist,hiroi} 
(although strictly speaking
a sharp distinction between $d$- and $s$-wave pairs does not exist in
ladders). While the main reason for the hole attraction is 
the minimization of broken rung singlets~\cite{dago1}, the appearance 
of a $d$-wave
character in the pairing is likely caused by the presence of short-range
antiferromagnetic fluctuations, establishing an interesting analogy 
between planes and ladders.

Recent work in this context using
the Density Matrix Renormalization Group (DMRG)~\cite{white} 
approach has confirmed
the presence of two hole bound states in the $d$-channel
in the case of 2-leg ladders~\cite{doug4}. However, in this study
it was remarked that holes in
the bound state spend part of the time located 
along the diagonals of the elementary
plaquettes in the problem, a feature somewhat counter-intuitive for
$d$-wave pairs since a two-body problem with $d_{x^2 - y^2}$
character cannot have particles in such a configuration.
In addition, the presence of strong spin singlets in the vicinity of
holes, resembling the Resonating Valence Bond (RVB) states characteristics 
of spin liquids~\cite{anderson},  were noticed.
In ladders an explanation for these results was
proposed in Ref.~\cite{doug4}: the singlets along plaquette diagonals have 
frustrating character and pairing occurs to share frustration.
As a consequence of this effect domain walls are induced in the problem.
A similar numerical result in the context of two-dimensional planes
was earlier reported using Exact Diagonalization (ED) techniques on
small clusters~\cite{didier2}.

The above mentioned recent
numerical studies motivated in part the present paper. Its purpose
is to further analyze the two hole problem
in planes and 2-leg ladders using Exact Diagonalization~\cite{review} 
and Truncated Lanczos~\cite{trunca} algorithms, 
as well as to propose intuitive arguments to
explain the results observed in
the present and previous investigations. 
The physics obtained with the DMRG method on ladders~\cite{doug4} 
is  found to be contained also in small clusters that can be handled
exactly.
An intuitive picture both in real and momentum space
is provided to justify how holes
can exist in the same sublattice in a 
$d_{x^2 - y^2}$-wave bound state. The importance of $retardation$
effects for this feature is remarked, and thus it is unavoidable to
conclude that time-dependent hole-hole 
potentials are needed to quantitatively account for the physics of the 
$t-J$ model. However, effective models with instantaneous interactions
are expected to
capture most of the relevant physics, specially regarding the symmetry
of the condensate, similarly as the BCS model does for the
electron-phonon problem.
The existence of strong spin singlets in the immediate vicinity of
tightly bounded
holes on planes
is here argued to be caused, at least in part, by 
the $d$-character of the bound
state and its existence is independent of the special features of the spin
background. In particular they exist in the case of an Ising background,
which does not have quantum fluctuations.
For the case of ladders, since spin singlets are
already formed in the undoped system our proposed explanation only
addresses the unexpected strong strength of some particular spin singlets,
and it is thus complementary to previous explanations~\cite{doug4}.
It is also reported here that
the special configuration where holes are located at short distances,
such as $\sqrt{2}$
lattice spacings, is just one of several equally important  hole
distances in the two hole problem, and in addition it becomes unstable
after the introduction of a realistic nearest-neighbor (NN) Coulombic
repulsion. 

\section{Holes in Same Sublattice in bound states 
transforming as ${x^2 - y^2}$}

\subsection{Discussion in Real Space}

As explained in the Introduction, previous 
studies~\cite{didier2,doug4} have emphasized
that the configuration where the two holes are located 
 in the same sublattice across the diagonals of a plaquette
has a substantial weight in the two-hole ground state.
At first sight this result
seems counter-intuitive since one is
used to the naive notion that in the $d_{x^2 - y^2}$ subspace
the wave function of two particles has a node along the lattice
diagonals. However, remember that this
 picture is based on the idealization of the two-hole problem
in a spin background (which is a $N-2$ body problem, where $N$
is the number of sites of the cluster under consideration)
as a system of only two particles in an otherwise
empty lattice interacting through a static effective potential. 
This potential is staggered in real space (peaked at
$(\pi,\pi)$ in momentum space), favoring the location of holes in
opposite sublattices. 
A wide variety of calculations including diagrammatic techniques
supplemented by Quantum Monte Carlo studies~\cite{doug2},
$t-J$ model estimations~\cite{review}, and other
approaches~\cite{dotsenko} all agree in this respect.
Using such a hole-hole potential
there is no doubt that the two particles
$must$ be located 
in opposite sublattices for the $d_{x^2 - y^2}$ symmetry to be
realized. Even within the context of the full $t-J$ model, if it were possible
to ``integrate out'' the
spin degrees of freedom, the resulting potential between the two holes
at  zero frequency must again induce the same feature.

For the case of planes, the key ingredient to understand
the apparent discrepancy between the numerical and
analytical calculations detailed above are the
$retardation$ effects caused by the finite velocity propagation of the
spin excitations in lightly doped antiferromagnets.
As a simple example consider a couple of holes located next to
each other in a perfect N\'eel 
spin background, as shown in Figs.1a-b. To construct a $d_{x^2 - y^2}$
wave function in this context, it is necessary to combine
Fig.1a and 1b with a negative relative sign. This provides
 the simplest example in the two hole subspace of a contribution to 
the $d_{x^2 - y^2}$ bound state. If the N\'eel state is considered
as the reference ``vacuum'', then each hole at site $i$
can be formally assigned an effective
spin according to the definition 
$S^{eff}_i = S_i - S^{AF}_i$, where $S_i$
is zero for a hole and $S^{AF}_i$ is the spin of the reference state at
the same site. With this definition holes at distance of one
lattice spacing have opposite effective spins.
However, there are other contributions to the many-body ground state of
two-holes in the $d_{x^2 - y^2}$ subspace caused by the presence of a
 nontrivial spin background.
For example, Fig.1c illustrates what
occurs if a hole is moved by one lattice spacing from its position in
Fig.1a. 
In this situation the
spin that is moved in the opposite direction of the moving hole
 is now located in 
the wrong sublattice with respect to the background. 
Once again if the N\'eel state is defined as the reference
vacuum, such an arrangement of holes
and spins is represented as in Fig.1d showing that it corresponds
now to a $three-body$ problem. The holes being now in the same
sublattice have the same spin projection according to the definition
of the spin $S^{eff}_i$ given above, and the overall
spin conservation is
achieved by realizing that the actual spin involved in the hole hopping
 process,
which now lives in the wrong sublattice, plays the role of 
a spin $one$ excitation with respect to the N\'eel background. Thus,
the hole hopping in the lightly doped $t-J$ model can be rephrased 
as the destruction of a hole with spin $\sigma$, and the creation
of a hole with spin $-\sigma$ and a bosonic excitation of spin $2\sigma$
(Note that the pair operators introduced in Ref.\cite{didier2}
also contains the notion that spin excitations are needed to have
$d$-wave symmetry in this context).
Formally this corresponds to a hopping process regulated by the
Hamiltonian

\begin{eqnarray}
H_{hopp} \propto \sum_{{\bf i}, \mu = \pm {\hat x}, \pm {\hat y}, \sigma} 
( c^\dagger_{{{\bf i}+\mu}, -\sigma}  c_{{\bf i}, \sigma }  a^\dagger_{{\bf i}, 2\sigma}
+ h.c. )
\end{eqnarray}

\noindent where the $c$-operators represent the (fermionic) holes, and the
$a$-operators the (bosonic) spin one excitations. The rest of the
 notation is standard.
Once the three-body character of the problem is understood,
it is easy to find combinations of individual states that transform
globally as
${x^2 - y^2}$. For instance, Fig.1e contains four states with 
holes located along the diagonal of a plaquette coexisting with a 
 spin excitation as described before.
If the four states have equal absolute
weight but the relative signs indicated in 
Fig.1.e, then it can be shown that 
a $d_{x^2 - y^2}$ state is formed. This reasoning has been tested by
solving a $3 \times 3$ cluster exactly fixing one hole at the center and
allowing for a second one to orbit around. Simply considering the
dominant configurations in the ground state, the three-body
characteristics of the motion of two holes around each other become clear.
In addition, certainly the discussion based on Figs.1a-e can be extended to
distances larger than those involved in a single plaquette.
Actually  as $J/t$ is reduced in the two hole problem, longer and
longer distances between the holes will have larger weights in the
wave function since $J/t$ regulates the strength of the attraction.

Based on this reasoning
the computational  results~\cite{didier2} reporting that
in planes
the configuration with holes located in the same sublattice 
has the largest weight in the two hole ground state can be
rephrased as an indication that 
the spin excitations are not ``instantaneous'' from the point of view
of the holes, but instead they have a finite lifetime. In other words,
the  process of holes moving within a given sublattice is not a rapid
tunneling process.
``Retardation'' effects apparently are important in models 
where antiferromagnetic correlations at
short distances are robust. This is not too surprising considering
that a typical spin wave velocity is regulated by the exchange $J$ while
a variety of numerical results have shown that the quasiparticle bandwidths
in the  $t-J$ and related models for cuprates are also of order $J$, at
least in the underdoped regime~\cite{review}.

\begin{figure}
\psfig{figure=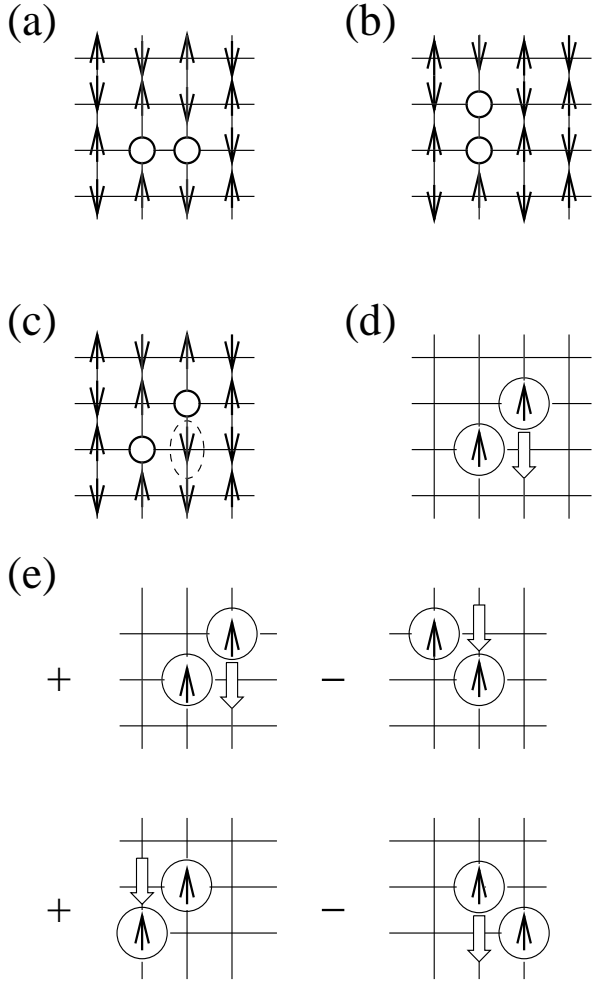,height=13.0cm,width=7.8cm,angle=-90}
\vspace{0.3cm}
\caption{
(a-b) Pictorial representation of a couple of holes in a spin N\'eel
state; (c) The movement of a hole in (a) introduces a spin
incorrectly aligned with respect to the staggered order in the 
background; (d) Taking the N\'eel state as reference (c) corresponds to
a three body problem, namely two holes and a spin excitation; (e) The
three bodies of (d) can be arranged in a $d_{x^2 - y^2}$ state as indicated.
}
\label{fig1}
\end{figure}

Note that the effect presented here is ``local'' in the sense that
it involves just a few sites of the lattice. Thus, naming the spin
excitation of Fig.1c as a spin-wave, which is usually associated to a
delocalized object, is somewhat misleading since
a localized excitation can actually be decomposed into planes waves
carrying a similar weight for all momenta.
Since the density of states of magnons peaks at ${\bf k} =
(\pi/2,\pi/2)$, these states may matter more for
local processes than the low-energy extended excitations of
momentum $(\pi,\pi)$. This is similar to what occurs in electron-phonon
problems where the phonons at the Debye frequency are more relevant than
the low-energy acoustic modes.
Then,  the spin-wave velocity (which is defined in
the vicinity of $(\pi,\pi)$ for spin problems)
is likely not the most important quantity to judge effects of
retardation in the formation of tight hole pairs.

\subsection {Strong spin singlets along plaquette diagonals:
an intuitive explanation}

The DMRG studies of Ref.\cite{doug4} 
for the case of two holes located at distance $\sqrt{2}$ also reported the
existence of a strong spin-singlet along the diagonal the opposite
to the one where the holes are located (for other recent studies 
see Ref.\cite{sierra}).  In the context of ladders their existence
was argued to be caused by the presence of 
frustrating singlet effects~\cite{doug4}. In
addition, singlets are already formed in the ground state of 2-leg ladders
and, thus, it is natural that they contribute to the movement of holes.
However, the strong strength of these ``diagonal'' singlets is unusual, and
even more strange is that they also appear in two-dimensional clusters as
shown numerically below. 
Then, the authors believe that there must be some other
ingredient contributing to the existence of these strong
diagonal singlets.
The discussion of the previous
subsection allow us to propose an explanation based on holes in
 an antiferromagnetic
background that complements previous discussions~\cite{doug4} based on spin
liquid states. Consider
once again Figs.1a-b remembering that these two states enter in the
wave function of a $d_{x^2 - y^2}$ state with opposite signs. Moving
the right hole in Fig.1a up one lattice spacing, Fig.1c is obtained
where the spins along the diagonal opposite to the holes have opposite
$z$-projections. This is once again reproduced in Fig.2a.
Now consider Fig.1b and move the upper hole to the
right one lattice spacing: in this case (Fig.2b) the spins along the
diagonal opposite to the holes are again antiparallel, but with projections
spin inverted with respect to those found in Fig.2a. Next, 
combine Fig.2a and
2b taking into account the fact that Figs.1a-b have a weight of opposite
sign in the ground state, and that the spins not belonging to the
plaquette being analyzed have not changed.
Then, it is clear that a $spin$ $singlet$ 
along the diagonal opposite to the holes is formed, as illustrated in
Fig.2c. Note that this reasoning should work specially well
 in an Ising background
(i.e. using a N\'eel state without quantum fluctuations), and in Sec.III
numerical results confirm this prediction using  the 
$t-J_z$ model~\cite{fye}.
Note also that for the case of an extended $s$-wave, the same line of
arguments would lead to a $triplet$ along the diagonal, a prediction
that will be tested numerically below.

A similar reasoning helps in understanding why singlets are formed
next to hole pairs even if they are not along diagonals. Consider Fig.1b
and move the upper hole first to the right and then down one lattice
spacing. The resulting spin configuration  is shown
in Fig.2d. Combining this state with Fig.1a and, again, remembering that
Figs.1a and 1b had a weight of opposite sign, now Fig.2e is obtained.
Then, the presence of singlets next to holes is also natural when
$d_{x^2 - y^2}$ states are considered, and they are there regardless of the
overall ground state properties of the spin system.

\begin{figure}
\psfig{figure=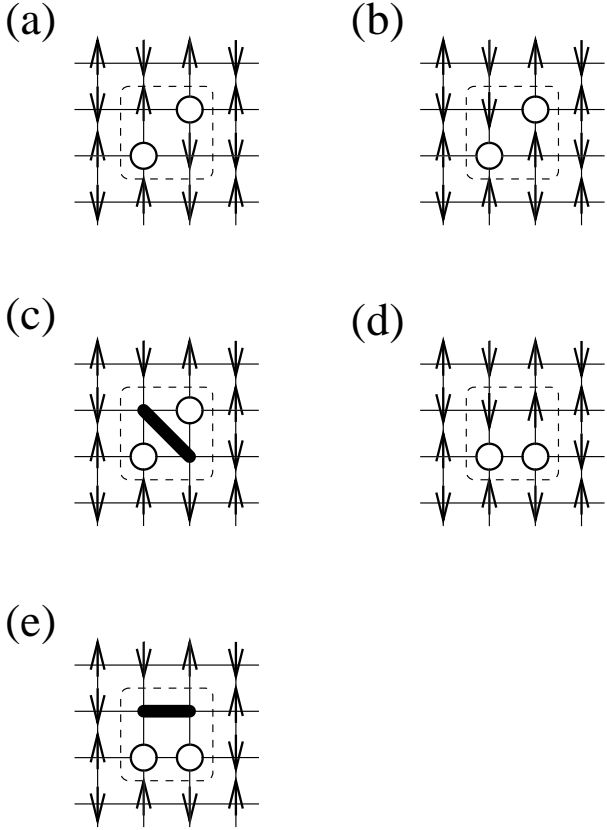,height=11.0cm,width=8.0cm,angle=-90}
\vspace{0.3cm}
\caption{
Spin and hole configurations allowing us to show that strong
spin singlets are expected in the vicinity of holes orbiting in a
$d_{x^2 - y^2}$ state when a N\'eel background is considered. 
For details see the text.
}
\label{fig2}
\end{figure}

\subsection{ Discussion in momentum space}

The previous discussion carried out in real-space in the limit
where holes are heavy  can be reinterpreted using a diagrammatic
approach in momentum-space considering fermions (holes) interacting
through the exchange of bosons (spin excitations). Fig.3a contains
a `` randomly'' selected diagram contributing to the bound state of
two holes in a spin background. The important point to remark arises 
if a snapshot of the system at
an arbitrary time (such as the one indicated by the dashed line) 
is taken: in this case it is clear that together
with the pair of fermions there may be a nonzero number of spin excitations.
In spite of this increase in the number of particles at intermediate
times, all must  be combined in such a way that the original 
spin and symmetry under spatial rotations of the fermionic pair at time
$t= -\infty$ is conserved.
In the case of relevance for the two-hole bound 
state the total momentum and spin
must remain zero, and the
$d_{x^2 - y^2}$ symmetry under
rotations must be preserved. Consider the intermediate time
denoted by the dashed line in Fig.3a:
  in this situation one vertex interaction
has taken place at previous times, changing the spin of the fermion
involved, and now
both fermions have the same spin projection with the boson
carrying the compensating spin, as in the real-space picture 
of Fig.1d. In addition, if the boson has momentum
$\bf q$,
the holes originally in $\bf -k$ and $\bf k$, now switch to $\bf -k$ and
$\bf k-q$. 
At this intermediate time and if $only$ the holes are considered
one may arrive to the counter-intuitive
 conclusion that they have switched to a
state of spin 1 and 
momentum $q$, that in addition is not in the $d_{x^2 - y^2}$ 
subspace (in principle they can be in any irreducible representations of
the $C_{4v}$ symmetry group of the square lattice). 
Such a conclusion would be incorrect since the overall quantum
numbers are given by the combination of fermions and bosons.
Concentrating on the behavior of only
the fermions in the problem  may lead to apparent paradoxes,
as described before.

\begin{figure}
\psfig{figure=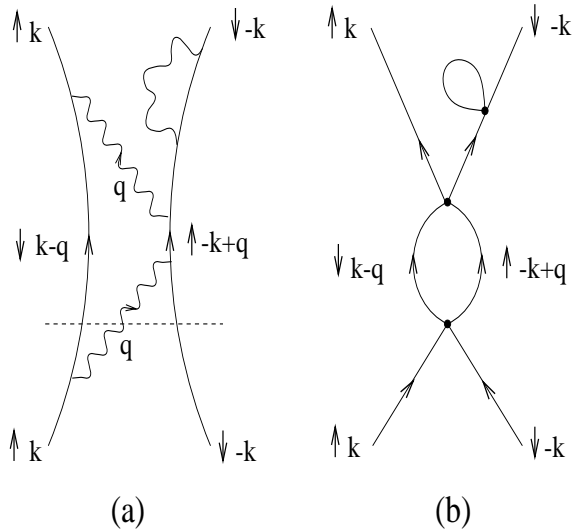,height=7.0cm,width=7.5cm,angle=-90}
\vspace{0.1cm}
\caption{
(a) Momentum space representation of the interaction of two holes
(solid lines) mediated by spin-waves (wavy lines). Spin and momenta are
indicated. The dashed line represents an intermediate time relevant for
the discussion in the text. Time runs upwards in the graph; (b) The same
Feynman diagram as in (a) but now assuming that the interchange of
spin-waves is instantaneous (no retardation). This collapses the
spin-wave propagators into a point. Momenta and spin are indicated.
}
\label{fig3}
\end{figure}

If the bosonic mediated interaction is considered instantaneous,
i.e. when a $\omega=0$ potential is used to describe the hole-hole
interaction, then a situation where at all times the holes 
carry the  quantum numbers they had at $t= -\infty$ is recovered 
(see Fig.3b).  In this case, holes must necessarily reside 
on opposite sublattices for its bound state
to transform as $x^2 - y^2$. Then, it is 
the fact that the $\omega \neq 0$ effects (retardation)
cannot be neglected in the $t-J$ model
 that allows for hole configurations belonging to the
same sublattice.
The diagrams in
 Figs.3a and 3b are similar to those arising in the context of
low temperature superconductors where electron-phonon interactions lead
to pairing. Fig.3a is the analog of an interaction mediated by
a phonon propagating slowly relative to the fermions,
 at a velocity regulated by the Debye
frequency, while Fig.3b plays the role of a diagram using only 
 the BCS model where the effective interaction is assumed to
be instantaneous and attractive. Note that in principle this is a drastic
approximation since the ratio of typical fermionic to phononic
velocities is about 100 in the low temperature superconductors.
Nevertheless, in this context it is 
known that the BCS model captures the essence
of the electron-phonon interaction, in particular the $s$-wave
symmetry of the superconducting condensate. Thus, it is  reasonable
to believe that a similar situation will occur in  models for the
Cu-oxides based on the interaction between holes and spin-waves where 
the relevant ratio of velocities is likely of order 1 (since both are
dominated by $J$ in the low hole density limit, as explained before).
In other words, assuming that the interaction mediated by the
spin-excitations is instantaneous is an approximation that 
preserves the relevant qualitative features 
of the problem, as the BCS model does in the electron-phonon problem.
In this respect is that models of
high-Tc such as those proposed in Ref.\cite{afvh}, where 
it is claimed that a peak
in the density of states of strong correlation origin causes the
appearance of an ``optimal'' doping, can be qualitatively correct
and useful as a starting point to understand spin-wave mediated
superconductivity. However, for a quantitative analysis the full
spin-hole problem with retardation effects included must be taken into
account, and that should be the main  message extracted from 
the numerical studies discussed here.


\subsection{Implications for SO(5) scenarios}

Note that the results of this section have implications for the
recently proposed SO(5) scenario for the cuprates~\cite{zhang}. Once again if a
snapshot of the system is taken at the
intermediate time (dashed line) of Fig.3a
the holes will be found to be in a triplet state of momentum ${\bf q}$, 
and this state will have low-energy.
However, as discussed before even though the lower energy spin-waves appear
at  $\bf q=Q$, their number is maximized at ${\bf q} \sim (\pi/2,\pi/2)$
and, thus, no particular value of the momentum is {\it a priori}
more important than others. This conclusion is in disagreement with the
results of Ref.\cite{zhang} that favor $\bf q = Q$ over other momenta.
In Fig.4, the energies of the two-hole ground state as a function of
momenta for the cases of a total spin zero and one and using the
 $t-J$ model are given. The results
are exactly calculated using a 18 site square cluster. In the
triplet branch there is no major difference between the different momenta,
in agreement with our discussion. The particular value $\bf q = Q$ is
not specially important in a broad range of couplings from $J/t = 0.2$
to 1.6. The present reasoning is simple and avoids the
use of complicated approximate symmetries of the $t-J$ and Hubbard 
models~\cite{greiter}.

\vspace{-0.3cm}
\begin{figure}
\psfig{figure=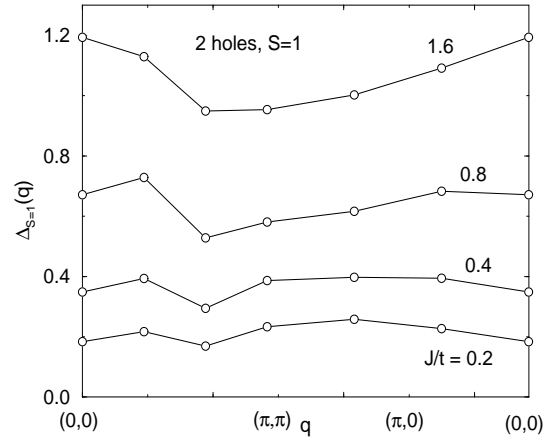,height=7.0cm,width=8.0cm,angle=-90}
\vspace{0.1cm}
\caption{
Energy $\Delta_{S=1}({\bf q})$ of the state with the lowest energy in the
subspace of two holes, momentum $\bf q$, spin one, and parametric with $J/t$.
The energies are referred to the overall  ground state energy of two
holes, which is a zero momentum spin singlet.
The results are
obtained exactly using an 18 site cluster. Results for 20 sites are very
similar.
}
\label{fig4}
\end{figure}

\section{Exact Diagonalization}

\subsection{Ladders}

Here it is addressed numerically
whether the recent DMRG results~\cite{doug4} for the behavior of two-holes in
the ladder $t-J$ model can be reproduced with other techniques. In 
particular, it would be desirable to obtain similar results in
clusters accessible to Exact Diagonalization (ED)
 methods since the implementation of
this algorithm is simple, calculations on a variety of
$t-J$-like models can be carried out without much effort,
periodic boundary conditions can be implemented,
and dynamical properties can be analyzed.

Using the ED method let us first investigate the
relative distance between holes in the case where two-hole bound states
are expected.
Fig.5 shows exact results obtained on
a $2 \times 10$ cluster with 2 holes and periodic boundary
conditions. The probability $P(d)$ of finding
the holes at a relative distance $d$ is presented against
 the coupling $J/t$. At every coupling, results are normalized such that
$\sum_d P(d) =1$. It is
clear from the figure that for realistic values of $J/t$, 
such as 0.3 or 0.4, the
highest chances indeed occur when $d = \sqrt{2}$ is the hole separation.
Thus, DMRG and Lanczos techniques provide similar
results regarding this issue which is gratifying. However, Fig.5
provides extra interesting information: $P(\sqrt{2})$ 
is actually similar to $P(1)$ and, thus,
results in this context must be necessarily interpreted in
 a probabilistic
sense  i.e. the distance with the highest chances is not necessarily
much relevant for the problem. For instance,
note that $P(\sqrt{2}) \approx 0.28$ which is substantially smaller than 1.

\vspace{-1.0cm}
\begin{figure}
\psfig{figure=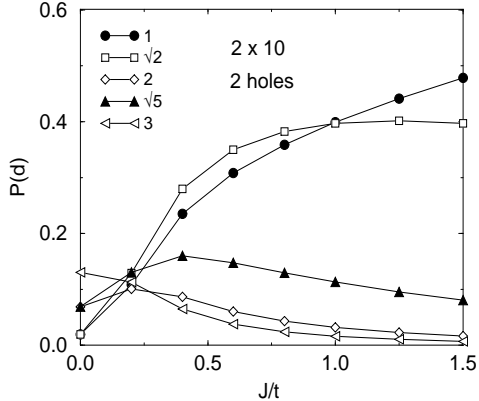,height=7.0cm,width=8.0cm,angle=-90}
\vspace{0.1cm}
\caption{
Exact Diagonalization results using a $2 \times 10$ cluster with two holes.
$P(d)$ is the probability of finding the holes at a distance $d$ apart
(convention indicated in the figure).
Results for several couplings $J/t$ are shown.
}
\label{fig5}
\end{figure}

This is better illustrated in Fig.6 where $P(d)$
at a fixed $J/t$ is provided once one of the holes is fixed to
an arbitrarily selected site of the 2-leg ladder. 
Working at $J/t=0.8$ the second hole is mostly
located 
on the chain the opposite to the first hole, more specifically
in the three sites the closest to it. As $J/t$ is
reduced to 0.4, the second hole spreads further its wave function
and the largest probability is at distance $\sqrt{2}$. However,
visually it is clear that a better representation is to imagine
the hole as moving  freely within a small region in the
vicinity of the fixed hole, rather than assign special importance
to one of the possible distances in this region.
In addition, note that in the results of Fig.5, and considering
data for other couplings not
shown explicitly, no
abrupt changes were observed as a function of $J/t$ for the
hole distribution and smoothly
the size of the hole pair wave function grows as $J/t$ is reduced.
 Finally
at $J/t=0.2$ the second hole seems delocalized on the whole
ladder (suggesting the absence of binding, or a bound state of size
larger than the cluster considered here).

The DMRG studies~\cite{doug4} isolating the configuration of holes where 
they are located along the diagonal of a plaquette 
have also shown
the existence of a strong spin singlet across the opposite diagonal of
the same plaquette. A possible contribution to this
result based on the $d$-character of the two-hole state
 was provided in Sec.II.
As in the previous paragraphs,
first it would be interesting to investigate if similar
numerical information is obtained using ED techniques. In Fig.7 results
are presented once again on a $2 \times 10$ cluster and with two 
holes. The bonds
where the spin correlation 
$\langle { {{\bf S}_{\bf i} }\cdot{ {\bf S}_{\bf j}} }
\rangle$
has changed substantially compared with the undoped case are
highlighted, following a convention similar as used in Ref.\cite{doug4}.
At large $J/t$, and isolating the configuration where
the two holes are in the same rung (since it has the highest chances)
the spins along the
nearest-neighbor rungs form a  stronger spin singlet than in the 
undoped ladder.
At $J/t=0.4$, and now with two holes in the $\sqrt{2}$-configuration,
the ED results confirm the previous DMRG calculations since indeed a robust
spin singlet is found in the same plaquette where the holes are located.
The considerable
strength of the diagonal spin singlet naturally
weakens the bonds that
link the plaquette being analyzed with the rest of the ladder. 
If $J/t$ is further reduced to 0.2, now the most likely
hole configuration have the holes at distance $\sqrt{5}$.
Concentrating on such configuration,
it is interesting to notice that here there are two
spin singlets along the plaquette diagonals (see Fig.7). Both are strong,
causing the bonds between them to be weaker than in the undoped case. 
However, appealing as this picture may
seem, the results in Fig.6 suggested that the 
holes are basically unbounded for this coupling. 
Thus, extracting conclusions out of snapshots of hole configurations 
is somewhat risky.

\begin{figure}
\epsfxsize=8.0cm
\hspace{0.3cm}\epsffile{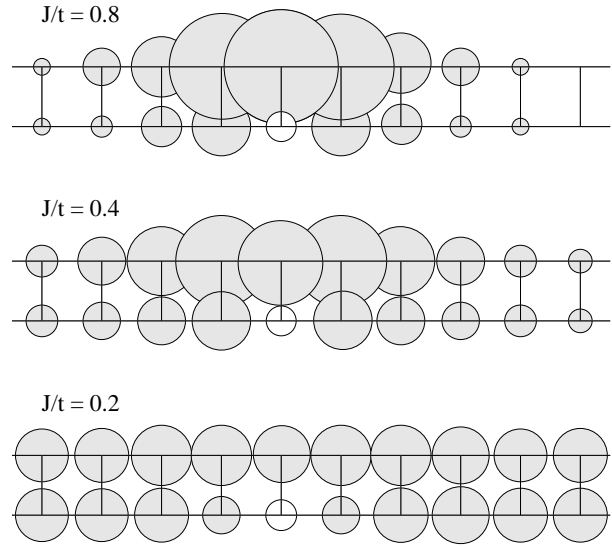}
\vspace{0.3cm}
\caption{
Ground state results obtained on a $2 \times 10$ cluster with two holes
solved exactly. One hole is fixed at the position denoted by the open
circle. The area of the gray circles is proportional to the probability
of finding the other hole at a particular site. Results for several
couplings are shown.
}
\label{fig6}
\end{figure}

As a partial summary, in this section it was found
that 
(i) ED techniques reproduce the pattern of
spin singlets found previously in DMRG studies once the position of
holes are fixed,  but (ii) 
several other hole configurations have comparable probabilities.
The two hole bound state for realistic values of $J/t$ has a finite size
and it resembles a bi-spin-polaron, with two particles moving
quasi-freely
inside a region of space regulated by the coupling.
There are no sharp
hole distances dominating the ground state of the problem.
The spins inside the two-hole wave function form strong
spin singlets which is natural since in the undoped case
the ground state has an RVB character. However, it was here
argued that the $d$-wave properties of the state contribute at least 
in part
to the strong spin singlet formation.

\begin{figure}
\epsfxsize=8.0cm
\epsffile{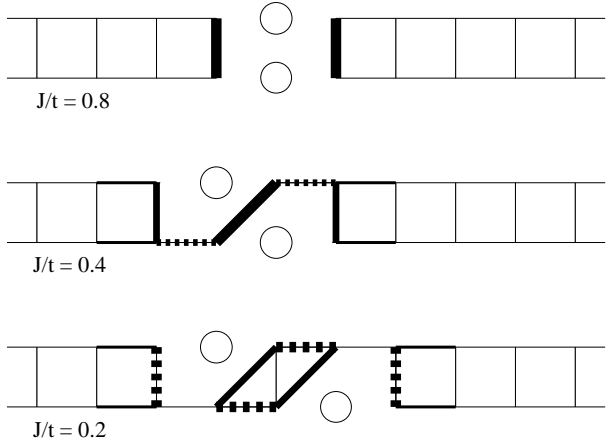}
\vspace{0.6cm}
\caption{
Results for two holes obtained at several couplings $J/t$
on a $2 \times 10$ cluster solved exactly. The holes are fixed at the
position with the highest chances in the ground state. Working in this
subspace, bonds where  the 
nearest-neighbor spin-spin  correlation has the largest
variation with respect to the undoped case are indicated.
Solid (dashed) bonds indicate correlations which are larger (smaller)
than in the undoped case by an amount
larger than 20\%. The thickness of the lines
is proportional to the change observed.
}
\label{fig7}
\end{figure}

\subsection{2D clusters} 

Calculations similar  as those in the previous subsection 
are here repeated using finite two-dimensional clusters. Fig.8
shows $P(d)$ for a $\sqrt{20} \times \sqrt{20}$ cluster 
solved exactly. 

\vspace{-0.8cm}
\begin{figure}
\psfig{figure=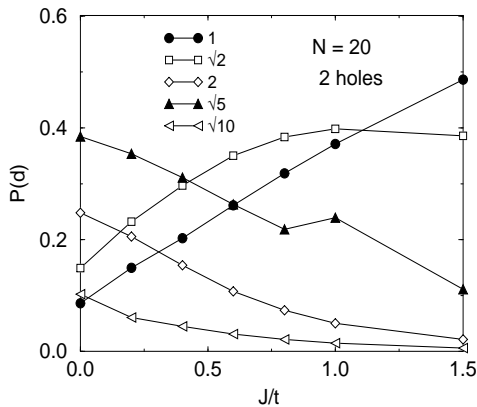,height=7.0cm,width=8.0cm,angle=-90}
\vspace{0.1cm}
\caption{
Exact Diagonalization results using a 20 sites square
cluster with two holes.
$P(d)$ is the probability of finding the holes at a distance $d$ apart
(convention indicated in the figure).
Results for several couplings $J/t$ are shown.
}
\label{fig8}
\end{figure}

As in the case
of the 2-leg ladders, a finite window in $J/t$ exists
where the distance $d=\sqrt{2}$ has the highest
chances, in agreement with Ref.\cite{didier2}. 
However, if $J/t=0.3$ or 0.4 is considered
realistic, then holes at $d=\sqrt{5}$ have
the largest probability.
However, as in the case of ladders there are several distances with
similar weight
i.e. $P(d)$ is not sharply peaked at one particular value of $d$.
Actually,
Fig.9 provides a pictorial
representation of the probability of having one hole at a given position
once the other hole is fixed at an arbitrary site, at three values of
 $J/t$.
As in the case of ladders, for large
$J/t$ the wave function of the second hole is localized in the immediate
vicinity of the first one. Reducing $J/t$ the extension of the
wave function smoothly increases, and at $J/t=0.2$ it covers the
whole cluster suggesting the absence of a bound state.
Thus, it seems that the bound state of two holes in 2-leg ladders and
planes share many similarities, at least in the regime of small and
intermediate size pairs. This universality may be caused by
the presence of robust antiferromagnetic correlations in both systems. 
The bound state
main features seem independent of the long distance  (gapped vs
gapless) characteristics of the undoped ground state.

The study of the pattern of spin singlets near holes shown in Sec.III.A
can be repeated for 
the 2D clusters as well, and results are given in
Fig.10. At large $J/t = 1.6$,
the holes are located with the highest chances at a distance of one
lattice spacing. As in the case of 2-leg ladders, the bonds parallel
to the location of the holes contain strong spin singlets, which weakens the
neighboring bonds along the same direction. When $J/t$ is
reduced eventually having the holes along a plaquette diagonal
becomes the configuration with the highest chances 
(with the caveats of the previous paragraphs).
In this case the spin singlet along the diagonal the 
opposite to where the holes are located is very strong, and as a
consequence the other bonds
associated with this spin singlet are weak, as it occurs for
ladders. At small $J/t=0.2$, once again two strong
spin singlets are found between the holes which themselves have the highest
chances of being located at distance
$\sqrt{5}$, again similarly as observed for the
same coupling on ladders.
 
A possible explanation for the presence of strong singlets was provided in
Sec.II where it was proposed that they
arise as a natural consequence of the $d_{x^2 - y^2}$
symmetry of the two-hole ground state in a N\'eel background. 
To verify this statement, ED studies for the spin anisotropic
$t-J$ model were carried out. In this case the Heisenberg
${ {{\bf S}_i}\cdot{{\bf S}_j} }$ interaction is replaced by
$S^z_i S^z_j + \lambda ( S^x_i S^x_j + S^y_i S^y_j)$. If $\lambda =1$
($\lambda =0$)  the Heisenberg (Ising) limit is recovered. Working
at $J/t = 0.6$ as an example, on 16 and 20 site 
clusters, and concentrating on two holes in 
the $d_{x^2 - y^2}$ subspace~\cite{comm90}
at distance $\sqrt{2}$,
the numerical results show that
as $\lambda$ is reduced from 1 to 0 the 
$\langle S^z_i S^z_j \rangle$ correlation (along the diagonal the opposite
to where the holes are) is negative, of similar value in this interval
of $\lambda$ (e.g. on the 20 site cluster, 
$\langle S^z_i S^z_j \rangle \sim -0.21$ 
at $\lambda = 0$ and $\sim -0.19$

\begin{figure}
\psfig{figure=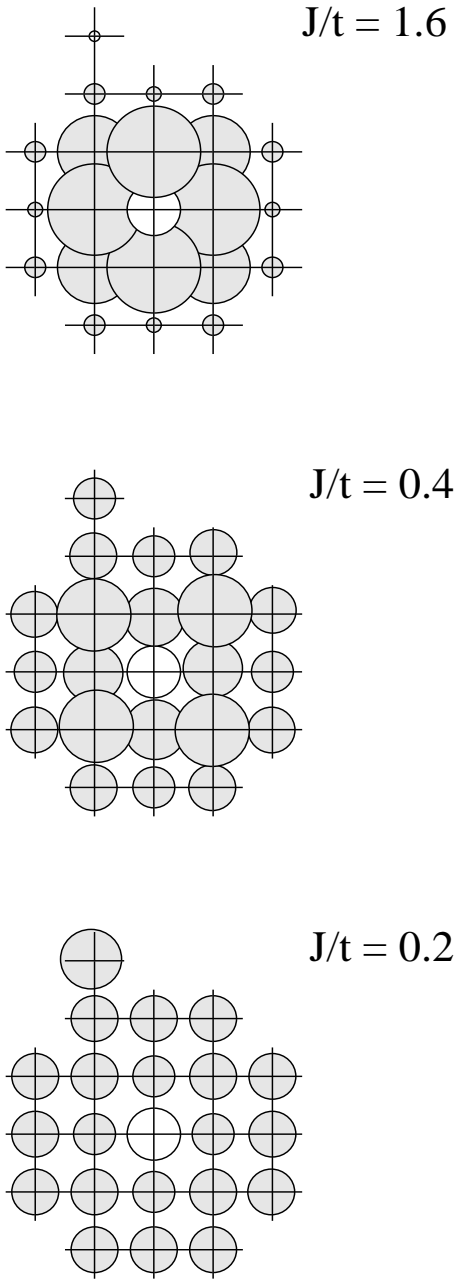,height=17.0cm,width=6.0cm}
\vspace{0.6cm}
\caption{
Ground state results obtained on a 20 sites square
cluster with two holes at $J/t = 1.6$, $0.4$, and $0.2$
solved exactly.
One hole is fixed at the position denoted by the open
circle. The area of the gray circles is proportional to the probability
of finding the other hole at a particular site.
More than 20 sites in the clusters are shown for clarity (periodic boundary
conditions were used).
}
\label{fig9}
\end{figure}

\noindent
at  $\lambda =1$). The value of the correlation changes smoothly as
a function
of $\lambda$ implying that at least in part the presence of strong
diagonal singlets in the Heisenberg limit is indeed caused by effects
contained in N\'eel backgrounds, as explained in Sec.II.
Very similar results have also been obtained for the $t-J_z$ model on a
$2 \times 8$ ladder: considering $J/t = 0.4$ as example, 
$\langle S^z_i S^z_j \rangle \sim -0.21 $ at $\lambda = 1.0$ 

\begin{figure}
\psfig{figure=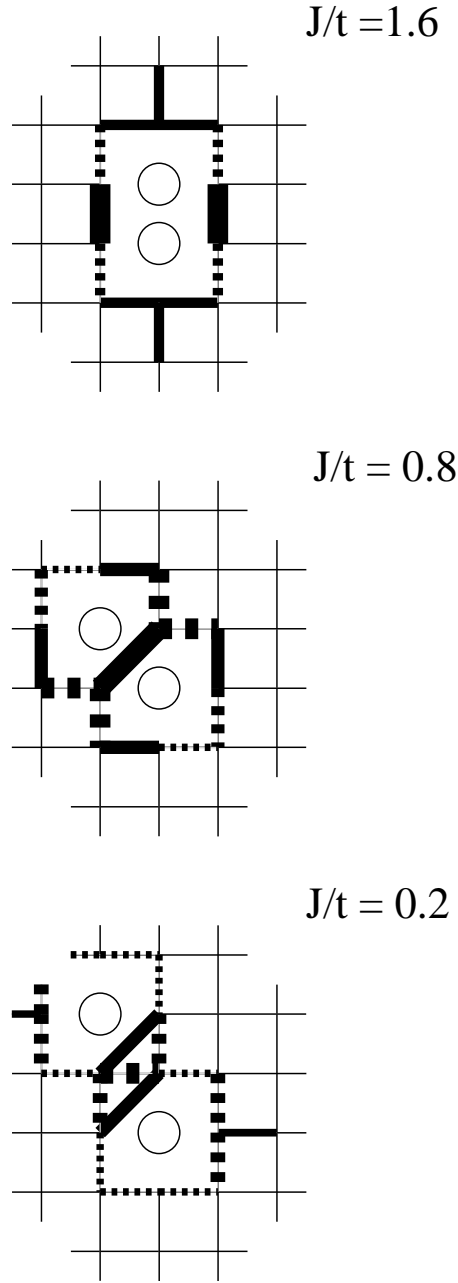,height=17.0cm,width=6.0cm}
\vspace{0.6cm}
\caption{
Results for two holes obtained at several couplings $J/t$
on a 20 sites square cluster using the Exact 
Diagonalization technique.
The holes are fixed at the
position with the highest chances in the ground state. Working in this
subspace, bonds where  the 
nearest-neighbor spin-spin  correlation has the largest
variation with respect to the undoped case are indicated.
Solid (dashed) bonds indicate correlations which are larger (smaller)
than in the undoped case by an amount
larger than 20\%. The thickness of the lines
is proportional to the change observed.
More than 20 sites in the clusters are shown for clarity (periodic boundary
conditions were used).
}
\label{fig10}
\end{figure}

\noindent
and
$\sim -0.22$ at $\lambda = 0.4$. Then, the effect appears also in
ladders and in this case it complements the tendency to form singlets
which is natural in this type of geometry.
In addition, similar studies for the case $\lambda =0$
have allowed us to verify that in the
subspace of $s$-wave states the diagonal singlet is now a $triplet$,
also in agreement with the predictions of Sec.II.
Then, this establishes that the antiferromagnetic fluctuations and
the $d$-wave character of the two-hole bound state contribute
significantly to the formation of the diagonal strong spin singlet
in ladders and planes.

\section{Truncated Lanczos}

It would be interesting to check if the results obtained in the previous
section survive an increase in the lattice size, specially for the case
of the two-dimensional clusters. {\it A priori} the strong similarities
between  the DMRG and ED results on small clusters for the
case of 2-leg ladders suggest that the two-hole bound state is not
much affected by size effects, at least for the case of tight pairs.
Nevertheless, even if only for completeness, the issue of size effects
is here addressed using 
the so-called ``Truncated'' Lanczos (TL) approach~\cite{trunca}. 
This technique is
currently under development and it has the advantage of sharing the
same good features of the ED method, specially (i) the ability of
producing
dynamical information, (ii) the generation of results using states with
momentum as a good quantum number (e.g. if periodic boundary
conditions are used), and (iii) the ability to
study Hamiltonians
that include interactions at intermediate and large distances if needed.
Thus, in addition to verifying the results previously discussed, 
this section has as an extra goal the test of the TL
method in a physically relevant case. 
Although it is doubtful that
the TL method will reach the same accuracy of the DMRG approach in the
study of static properties of quasi-one-dimensional systems, it may
provide an intermediate algorithm between ED and DMRG 
keeping some of the main features of both techniques. Actually, recent
developments~\cite{newbasis}
in this context using an exact change of basis to work with better degrees of
freedom for the truncation procedure have already produced promising results
that may transform the TL method into a more widely used technique for
the study of correlated electrons.

Since the details of the Truncated Lanczos
 approach were described before in the literature\cite{trunca},
here only a brief summary will be included. In the
first step of this
method one state of the full basis (belonging to the
subspace of momentum and total spin $z$-projection that will be investigated)
is selected to start the iterations. In the analysis below 
the zero momentum N\'eel
state was used. Note that this starting point is somewhat inefficient
for the case of ladder systems where the ground state has no long-range
order and a spin-gap,
and the recent developments~\cite{newbasis} mentioned before 
have indicated that a better starting point
would be to use rung spin singlets in the singlet-triplet basis. 
Nevertheless, for the particular case of two holes in the $t-J$ model 
the approach in the $S^z$-basis seems accurate (as shown below)
and, thus, results in the new basis will be postponed for a future publication.
After the initial state is chosen, the Hamiltonian is
applied several times producing a basis set of a few hundred
states. In this space, generated dynamically by the Hamiltonian,
a Lanczos diagonalization is performed.
The ground state wave function is analyzed and only the basis states with a
weight $|c|^2 > \lambda$ are kept. The cutoff $\lambda$ is selected
such that the number of states after the truncation procedure is performed
is only about 50\% or less of the size of the matrix being studied.
The procedure is
repeated several times i.e. growing the space, diagonalizing in the
generated basis, truncating to a fraction of it
(``back and forth'' procedure), until the 
available amount of
memory is exhausted.  This slow-growth approach 
has proven to work very well in
some cases such as the $t-J_z$ model~\cite{trunca}, 
and it is believed to produce
reasonable results in spin-gapped models. However, below it will be shown that 
it works also for two-dimensional clusters that have low-energy excitations.

\subsection{Truncated Lanczos on Ladders}

Fig.11 shows the ground state energy as a function of the size of
the basis for a $2 \times 16$ cluster with 2 holes. This particular
example can not be solved exactly with present day computers since
the total Hilbert space contains $\sim 10^9$ states, even after
translational invariance is used to reduce the basis size. However, Fig.11
suggests that  the

\vspace{-0.6cm}
\begin{figure}
\psfig{figure=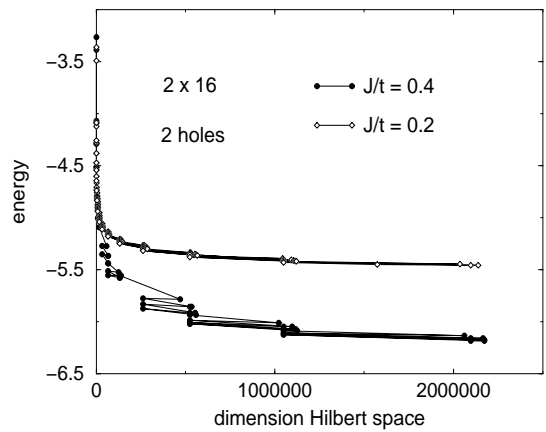,height=7.0cm,width=8.0cm,angle=-90}
\caption{
Ground state energy obtained using the Truncated Lanczos
approach working at $J/t=0.2$ and $0.4$, as a function of
 the size of the basis
used. The cluster has $2 \times 16$ sites and there are two holes.
The oscillations are produced by the ``back and forth'' procedure
described in the text. For both couplings the energy is measured
with respect to the energy of
the N\'eel state with two holes in the same rung, namely
$E = -J(48-5)/2$.
}
\label{fig11}
\end{figure}

\noindent
 TL approach with only $\sim 2 \times 10^6$ states
produces an energy 
converged up to 3 significant figures. The oscillations in the
energy shown in Fig.11 as the dimension of the Hilbert space grows
 are a consequence of the ``back and forth''
procedure described before.

In the absence of an exact value for the ground state energy of the
cluster considered in Fig.11, it
is possible to judge the accuracy of the approximate state obtained
using the TL method by evaluating correlation functions.
In  Fig.12 a calculation similar to that reported in Fig.7 is presented,
namely fixing $J/t$ the configuration of holes with the
highest chances is selected, and in that subspace the spin correlations
in the vicinity of the holes are shown. The links where the
 nearest-neighbor spin-spin correlation
functions have the largest variation with respect to the undoped system
(also calculated with the TL method) are highlighted.
Qualitatively the results are very similar to those found using
the $2 \times 10$ cluster, i.e. for $J/t=0.4$ the 
$\sqrt{2}$ hole configuration has the highest chances
 and the spin singlet along the
opposite diagonal
is strong, while for $J/t=0.2$ the holes are at distance $\sqrt{5}$ with two
spin singlets formed in between. The excellent agreement between the
results obtained with the  TL method compared with ED and DMRG techniques
is somewhat surprising 
since in the TL method applied to the $S^z$ basis
 the spin-spin correlations away from
the holes indicate correlations stronger than expected for a spin-gapped
system. However, in spite of this fact the behavior of holes is not affected,
reinforcing the notion that the physics of
tight hole pairs is dominated by the presence of spin correlations at
short distances independently of their long-range behavior.
Similar conclusions 

\begin{figure}
\epsfxsize=8.0cm
\epsffile{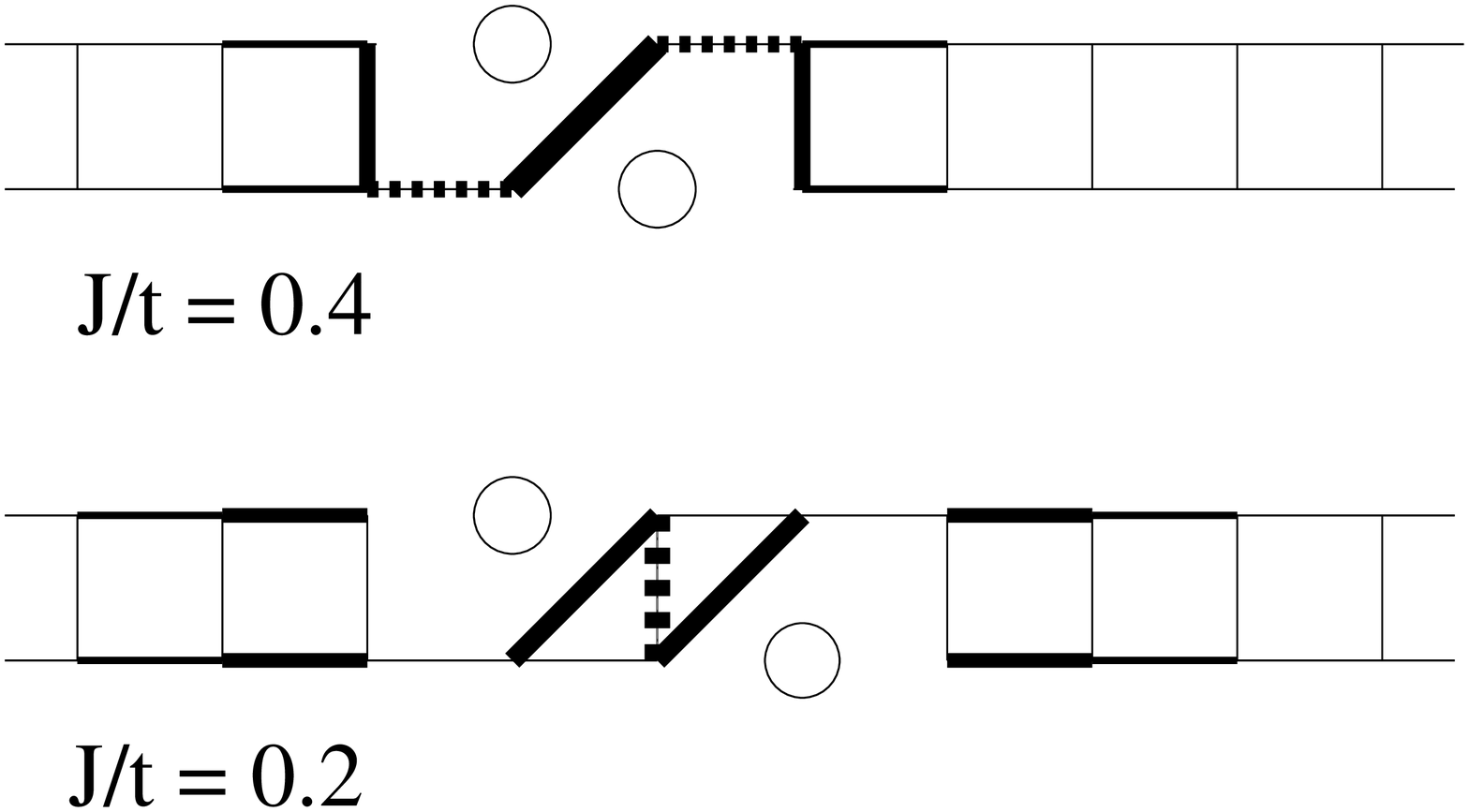}
\vspace{0.6cm}
\caption{
Results for two holes obtained at several couplings $J/t$
on a $2 \times 16$ cluster using the Truncated Lanzcos 
algorithm keeping $\sim 2 \times 10^6$ states.
The holes are fixed at the
position with the highest chances in the ground state. Working in this
subspace, bonds where  the 
nearest-neighbor spin-spin  correlation has the largest
variation with respect to the undoped case are indicated.
Solid (dashed) bonds indicate correlations which are larger (smaller)
than in the undoped case by an amount
larger than 20\%. The thickness of the lines
is proportional to the change observed.
}
\label{fig12}
\end{figure}

\noindent
can be reached analyzing other observables.
For instance, Fig.13 shows $P(d)$ for the $2 \times 16$ cluster, and
a couple of couplings. Several distances have comparable probabilities.
In particular for
$J/t=0.2$ there is an approximate plateau in $P(d)$
 between $d=1$ and $3$.

\vspace{-0.6cm}
\begin{figure}
\psfig{figure=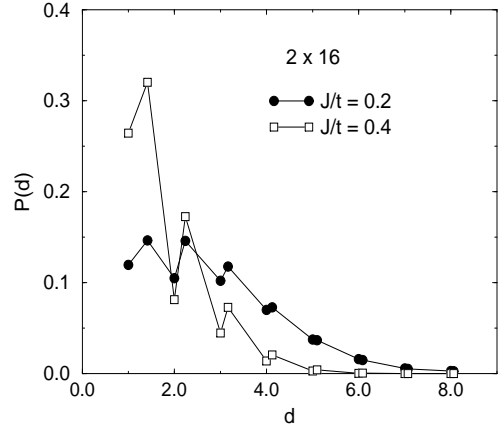,height=7.0cm,width=8.0cm,angle=-90}
\vspace{0.4cm}
\caption{
Probability $P(d)$ of finding the holes at a distance $d$
on a $2 \times 16$ cluster with two holes, calculated using the
Truncated Lanczos algorithm. Results for two couplings $J/t$ are
presented.
}
\label{fig13}
\end{figure}

\subsection{Truncated Lanczos on 2D Clusters}

Encouraged by these results, the TL approach was applied
to the 2D $\sqrt{32} \times \sqrt{32}$ cluster with 2 holes. Although the 
basis needed to solve exactly this problem can be reduced by a factor 4 
compared with
the basis used for the ladder (due to rotational invariance on square
clusters), the problem is difficult to solve exactly~\cite{leung}
and in addition it has not
been addressed with DMRG techniques.
Fig.14 contains TL results using $\sim 2 \times 10^6$ states isolating
the hole configuration with the highest chances and studying the spin
background
in its vicinity.
Once again, the results are almost identical to those
found using smaller clusters (Fig.10).
The dynamically generated strong plaquette-diagonal 
spin singlet clearly
appears in the 2D cluster study, similarly as it occurs in 2-leg ladders.
Fig.15 shows that the same agreement with smaller cluster calculations
occurs 
once the hole-hole
correlation $\langle n(0) n({\bf  r}) \rangle$, where $n({\bf r})$
is the hole number operator at site $\bf r$, 
is calculated. For $J/t = 0.4$ 
the correlation is maximized at distance $\sqrt{2}$. However,
remember that a given site has 4 neighbors at distance $\sqrt{2}$ but
8 at distance $\sqrt{5}$, and thus the maximum probability $P(d)$ is at
$d = \sqrt{5}$ in the case of $J/t = 0.2$. 
Fig.16 shows similar information in a
different representation. The results are in good qualitative agreement
with those found in smaller clusters.

\begin{figure}
\epsfxsize=8.0cm
\epsffile{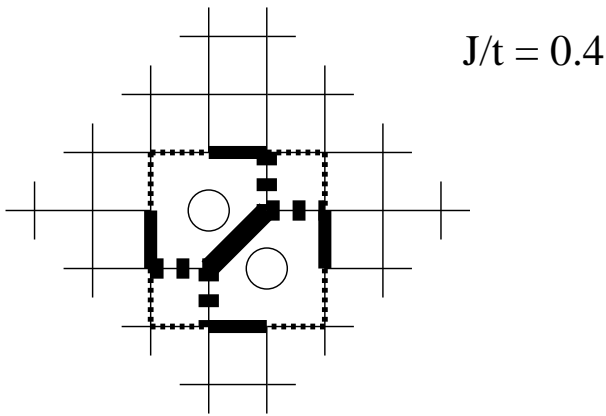}
\vspace{0.6cm}
\caption{
Results for two holes obtained at $J/t=0.4$
on a 32 sites square cluster  using the 
Truncated Lanczos algorithm keeping $\sim 2 \times 10^6$ states.
The holes are fixed at the
position with the highest chances in the ground state. Working in this
subspace, bonds where  the 
nearest-neighbor spin-spin  correlation has the largest
variation with respect to the undoped case are indicated.
Solid (dashed) bonds indicate correlations which are larger (smaller)
than in the undoped case by an amount
larger than 20\%. The thickness of the lines
is proportional to the change observed.
}
\label{fig14}
\end{figure}

\begin{figure}
\psfig{figure=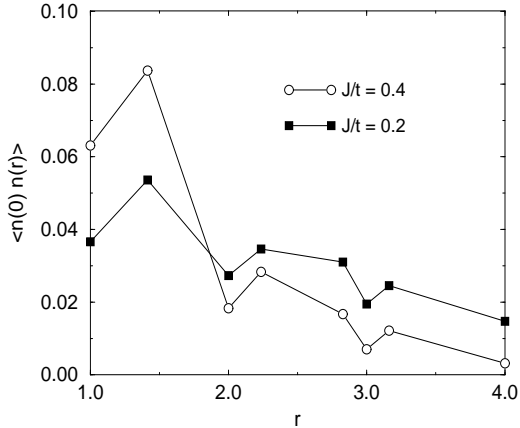,height=7.0cm,width=8.0cm,angle=-90}
\vspace{0.4cm}
\caption{
Hole-hole correlation $\langle n(0) n(r) \rangle$ vs distance $r$
obtained with the Truncated Lanczos method on 
a 32 sites square cluster 
and two holes keeping $\sim 2 \times 10^6$ 
states. The couplings are indicated. 
}
\label{fig15}
\end{figure}

\begin{figure}
\epsfxsize=8.0cm
\epsffile{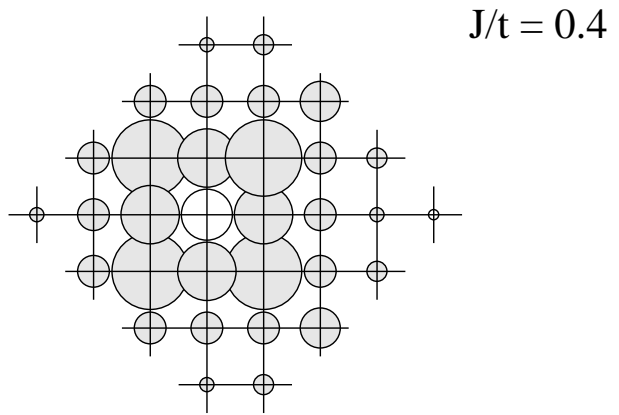}
\vspace{0.4cm}
\caption{
Ground state results obtained on a 32 sites square
cluster with two holes at $J/t=0.4$ using the Truncated Lanczos 
algorithm keeping $\sim 2 \times 10^6$ states.
One hole is fixed at the position denoted by the open circle. The 
area of the gray circles is proportional to the probability
of finding the other hole at a particular site.
}
\label{fig16}
\end{figure}

\section{Influence of hole-hole electrostatic repulsion}

Results such as those contained in Figs.6,9 and 16 clearly imply that
a pair of holes form a bound state on  2-leg ladders and planes in the
realistic regime of couplings close to $J/t \sim 0.4$, in agreement with
a vast amount of previous literature~\cite{review}.  At this
coupling the ``size'' of the pair (defined as the
mean distance between holes)
is about a couple of lattice spacings. Since for $J/t=0.2$, the bound state
seems to have disappeared (according again to the same Figures), then the
pair size must change very rapidly in the window of couplings
between 0.2 and 0.4, where a ``critical'' value $J/t|_c$ must exist
leading to hole binding.

Let us analyze in more detail a particular case such as $J/t=0.4$ which
is located in a window of couplings ``a priori'' presumed to be realistic.
Here holes are mainly located 
at distances 1, $\sqrt{2}$, and $\sqrt{5}$ lattice
spacings from each other. 
Such small distances naturally raise 
a couple of concerns: first, experimental results for YBCO have
suggested that its coherence length $\xi$ is about $15 \AA$ which 
roughly corresponds to 4 lattice spacings~\cite{cyrot}. 
Doped La-cuprates have an even higher $\xi$ of
about $35 \AA$. More recent estimations arrive to
$\xi = 18.3 \AA$ and $22.7 \AA$ for 1\%
Zn-doped underdoped YBCO and La-214, respectively~\cite{muon}.
These lengths are certainly much shorter than those
observed in low temperature superconductors, but still do not locate the
cuprates in the regime where $\xi$ is smaller than the mean distance
between carriers. If $\xi$ is interpreted, as usually done, as 
the size of the Cooper pairs in the superconducting condensate
and if the $t-J$ model is used,
then $J/t$ must be fine-tuned closer to  its critical value $J/t|_c$
to have hole pairs of size equal to the experimentally measured
 $\xi$. In other words, at $J/t = 0.4$ the hole pairs in the $t-J$ model are
substantially smaller in size than needed to represent the cuprates.
Then, from this point of view the peculiar properties of a very
tight bound
state are not much relevant, and only the qualitative aspects of the
problem can be extracted from studies of the pure $t-J$ model at $J/t=0.4$.

In addition, having pairs of small size raises concerns regarding the
stability of such pairs once the Coulomb interaction is considered.
While the on-site electronic repulsion is usually taken  into account in the
framework of strongly correlated electrons, the repulsion at the next
relevant distance of one lattice spacing is usually neglected (with an
exception being the analysis of stripe formation in the cuprates~\cite{tj3}).
Other authors have raised similar concerns~\cite{miha}.
A rapid estimation 
shows that the Coulombic electrostatic effect
 cannot be neglected since the bare potential
energy between two charges at one lattice spacing from each other is
 $V_{NN}=e^2/a=~3.8~eV$, where $a = 3.8 \AA$ as in the Cu-oxides compounds.
Certainly this repulsion is considerably larger than the 
attraction expected between
holes which is regulated by $J \sim 0.1 eV$.
This source of attraction would be destroyed by the bare
Coulomb interaction, unless retardation effects are important.

However, it is reasonable to 
consider a more optimistic scenario where the Coulomb repulsion at
distance of one lattice is influenced by other orbitals in the
Cu-ions, polarization of oxygen~\cite{ishihara}, and possible 
metallic screening in the
doped regime.
To gain some intuitive insight about the influence of these effects
let us first simply divide $V_{NN}$ by the dielectric
constant $\epsilon_{\infty}$ 
obtained experimentally for the high-Tc compounds (certainly
being aware that the result will be qualitative at best since
one is not supposed to use such a constant
for the short distance effects discussed in this section).
Although there is a large range of estimations in the literature,
a number near
$\epsilon_{\infty} \sim 30-40$ is
 a reasonable assumption~\cite{chen}. However, even considering this factor
the new estimation of the Coulomb repulsion 
still gives a number
competing with the order of magnitude of the
 attraction caused by antiferromagnetism,
namely  $V_{NN}=e^2/(\epsilon_{\infty} a)=~0.1~eV$. 
A variety of other calculations lead to qualitatively similar results. 
The 
Cu-O repulsion ($V_{pd}$) has been estimated  
to be $\sim 1.0 eV$~\cite{other}, i.e. smaller than the 
bare Coulomb interaction at distance $1.9 \AA$ roughly by a factor 8.
If this same factor is applied at distance of one lattice spacing
the Cu-Cu repulsion ($V_{dd}$) should be about 0.5 eV.
In addition, note that the binding energy
of a hole pair in two-dimensional clusters for $J/t = 0.4$
is only
$\Delta_B \sim 0.2 J$~\cite{review} i.e.
while the hole attraction is regulated by $J$ the numerical 
coefficient that takes into account
the fact that holes actually lose kinetic energy by forming the pair
magnifies even more the damaging effects of the Coulombic repulsion.

Then, electrostatic effects
will likely make unstable the very tight pairs observed
in the  computer simulations described before.
To analyze explicitly this effect in Fig.17 results for the
probability of finding the holes at a given distance $d$ are provided
for the case of a 2-leg ladder~\cite{barnes}.

\begin{figure}
\psfig{figure=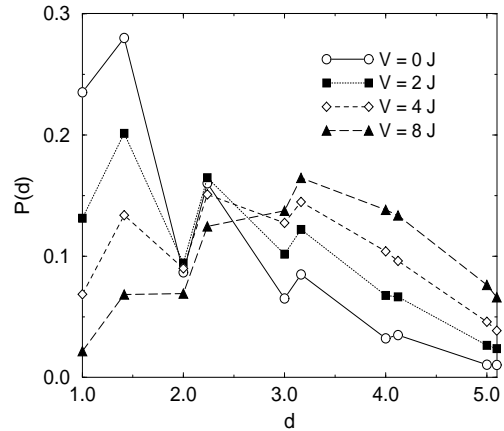,height=7.0cm,width=8.0cm,angle=-90}
\vspace{0.4cm}
\caption{
Probability $P(d)$ of finding the holes at a distance $d$
on a $2 \times 10$ cluster with two holes at $J/t = 0.4$, 
calculated using the Lanczos algorithm. 
Results for several values of the coupling $V$
regulating a nearest-neighbor repulsion between holes are
presented.
}
\label{fig17}
\end{figure}

\noindent
The Coulomb repulsion is included
adding to the $t-J$ Hamiltonian a term $V\sum_{\langle i j \rangle}
n_{\bf i} n_{\bf j}$, where $n_{\bf i}$ is the number operator at site
${\bf i}$ and the rest of the notation is standard. The results show
that the relevance of very small distances is actually lost
already when
 $V = 2J \sim 0.2 eV$. 
At larger values of the repulsion such as $V = 4J$ the
probability has a broad plateau between distances 2 and 3, 
and the bound state seem to
disappear when $V$ is increased slightly further~\cite{similar}.
Note, however, that if the hole-hole instantaneous Coulombic
 repulsion is assumed to be 
short-ranged more extended bound states may survive
the inclusion of such electrostatic energies since their size
increases as $J/t$ is reduced to its critical value. But this once again
will imply that a typical bound state size will be larger than just a
couple of lattice spacings.

There are two possible ways to avoid the problem of the Coulomb
interaction at short distances. One is by considering retardation
effects, and the other by finding other sources of screening that could
reduce drastically the previous estimations of $V_{NN}$. Let us consider
retardation first:
the discussion in the previous sections have indeed suggested that 
this effect  is important in the $t-J$ model and it may avoid the problem
of the Coulombic interaction similarly as it occurs in electron-phonon
systems, namely a hole could scramble the spin order in a region of
space during its
movement, and a second hole could take advantage of this distortion.
If one hole follows the other at a
distance of a few lattice spacings, then the short-distance Coulomb
repulsion can be avoided and the results would be compatible with
estimations of $\xi$ in the literature. As a simple illustration in
Fig.18, a hole creates a string 
excitation~\cite{boris} and a second hole heals the
damage in the spin background at a later time~\cite{hirsch}, but 
keeping some distance
to avoid repulsions. A similar process can potentially take place in the
case of ladders as also shown schematically in Fig.18. Considering the
ground state as dominated by rung singlets, the movement of a hole produces
the transformation of one of those singlets 
into a singlet along the diagonal of
a plaquette. Then, moving a hole introduces a damage in the spin 
configuration similar to what occurs with strings in the planes. The second
hole can heal this damage as shown in Fig.18.

\begin{figure}
\psfig{figure=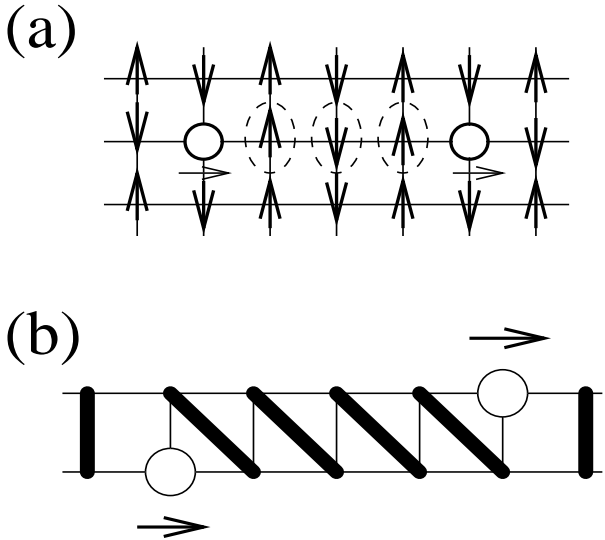,height=7.0cm,width=8.0cm,angle=-90}
\vspace{0.3cm}
\caption{
(a) Possible behavior of holes in an antiferromagnetic background
that avoids the short distance Coulombic repulsion. The first hole
moving, e.g., to the left creates a string of spins
 (highlighted in the figure) incorrectly aligned
with respect to the background. The second
hole can heal this damage following the first hole at some distance.
The picture is possible if the spin excitations have a large lifetime
such that the string of spins is not erased by spin fluctuations before
it is used by the second hole to improve its energy; (b)
Similar ideas as in (a) but now in the context of ladders. Here the
spin background is not antiferromagnetic but the spins mainly form 
singlets along the rungs. A hole moving along the chains creates a 
string of diagonal singlets. The second hole can heal this damage,
keeping some distance from the first to avoid the Coulombic repulsion.
}
\label{fig18}
\end{figure}

The second possible solution invokes a source of screening not
considered in the previous discussion. 
In the absence of other holes and using the $t-J$ model where
electron-hole pairs cannot be created, metallic screening effects 
should not affect the results and, thus, the Coulombic interaction
will actually be of long-range. However, in the presence of 
a finite density of holes, a simple Thomas-Fermi approximation can
be used to make a rough estimation of 
a possible metallic screening length $\lambda$. 
Considering 
a hole density $n \sim 7 \times 10^{21} cm^{-3}$~\cite{cyrot}, 
and using standard textbooks equations for three dimensional metals,
 $\lambda$ is found to be
$\sim 0.8 \AA$, which is small indeed and could potentially
drastically reduce the
effects of the Cu-Cu Coulomb interaction. However, it is clear that the
Cu-oxides are very different from the three dimensional metals where
Thomas-Fermi approximations are  qualitatively reliable. Thus,
the issue of whether metallic screening can reduce the effects of
Coulomb repulsions between holes in the cuprates remains an open question 
that deserves further studies.

\section{Conclusions}

In this paper the problem of two holes in a spin background with robust
antiferromagnetic correlations was
revisited. Using Exact Diagonalization and Truncated Lanczos techniques
applied to planes and ladders with up to 32 sites
the typical distances between holes was studied. It was observed that
the maximum probability occurs when the holes are at $\sqrt{2}$ lattice
spacings of each other, in agreement with previous calculations.
However, other distances were found to be equally 
relevant. An intuitive explanation
for these results in a real-space picture, as well as diagrammatically
was provided. Strong spin singlets are found near two holes in a
$d$-wave state. An explanation for this effect
based on a N\'eel background
 was presented, complementing other
approaches~\cite{doug4} based on spin disordered backgrounds.
It is concluded that holes in their movement create
spin excitations with a non-negligible lifetime, and thus $retardation$
effects are important in the $t-J$ model.
However, the instantaneous approximation captures properly the
qualitative aspects, specially the symmetry of the bound states which is
in the $d_{x^2 - y^2}$ channel.
The short size of the pairs raises concerns regarding the stability of
the bound states if $NN$  Coulomb interactions are included. Estimations of
the strength $V_{NN}$ of these interactions were 
here provided. A numerical study
showed that the bound states increase their size as $V_{NN}$ grows, and the 
typical hole distance in the pair can become comparable to the
experimentally measured coherence length, which is about 4 lattice
spacings, even if the values of $J/t$ used produced unrealistic small
pairs in the absence of the $NN$ repulsion.

\section{acknowledgments}

Conversations with  H.-B. Sch\"uttler, S. R. White, R. Eder, D. 
Poilblanc, and G. A. Sawatzky are gratefully acknowledged.
E.D. is supported by grant NSF-DMR-9520776. Additional 
support by the National High
Magnetic Field Lab and Martech has also been provided.


\begin{references}
\bibitem{similar} Note that similar calculations carried out in two
dimensional clusters that can be solved exactly are more severely
affected by size effects. Nevertheless the tendency to maximize the
distance between holes upon the introduction of the nearest-neighbor
Coulomb repulsion is clear.
\end{references}
\end{document}